\def\CLASSINPUTtoptextmargin{20mm}
\def\CLASSINPUTbottomtextmargin{27mm}
\documentclass[conference]{IEEEtran}

\usepackage{cite}
\usepackage{amsmath,amssymb,amsfonts}
\usepackage{amscd}
\usepackage{epsfig}
\usepackage{psfrag}
\usepackage{rotating}
\usepackage{float}
\usepackage{stfloats}
\usepackage{subcaption}
\usepackage{tikz}
\DeclareMathOperator{\Tr}{Tr}
\newcommand{\Real}{\mathrm{Re}}

\newcommand{\Pt}{P_t}
\usepackage{balance}
\usetikzlibrary{shapes.geometric}
\usetikzlibrary{arrows.meta}
\usepackage{pgfplots}
\usepackage{mathtools}
\usepackage{booktabs}
\usepackage{algpseudocode}
\usepackage{graphicx}
\usepackage{array}
\usepackage{siunitx}
\usepackage{algorithm2e}
\usepackage{float}
\DeclareSIUnit{\belmilliwatt}{Bm}
\DeclareSIUnit{\belsquaremeter}{Bsm}
\DeclareSIUnit{\bit}{bits}

\DeclareMathOperator*{\argmin}{arg\,min}
\graphicspath{ {./figures/} }

\usepackage{textcomp}
\usepackage{xcolor}
\usepackage[nolist,nohyperlinks]{acronym}
\pgfplotsset{compat=1.18}
\long\def\comment#1{}

\newfont{\bbb}{msbm10 scaled 700}

\newfont{\bb}{msbm10 scaled 1100}

\newcommand{\EE}{\mbox{\bb E}}

\newcommand{\av}{{\bf a}}

\newcommand{\fv}{{\bf f}}
\newcommand{\gv}{{\bf g}}
\newcommand{\hv}{{\bf h}}

\newcommand{\nv}{{\bf n}}

\newcommand{\sv}{{\bf s}}

\newcommand{\uv}{{\bf u}}
\newcommand{\wv}{{\bf w}}

\newcommand{\xv}{{\bf x}}
\newcommand{\yv}{{\bf y}}

\newcommand{\Cm}{{\bf C}}
\newcommand{\Dm}{{\bf D}}

\newcommand{\Hm}{{\bf H}}
\newcommand{\Id}{{\bf I}}
\newcommand{\Jm}{{\bf J}}

\newcommand{\Tm}{{\bf T}}

\newcommand{\etav}{\hbox{\boldmath$\eta$}}

\newcommand{\nuv}{\hbox{\boldmath$\nu$}}
\newcommand{\muv}{\hbox{\boldmath$\mu$}}

\newcommand{\tauv}{\hbox{\boldmath$\tau$}}

\newcommand{\diag}{{\hbox{diag}}}

\renewcommand{\arg}{{\hbox{arg}}}

\newcommand{\herm}{{\sf H}}
\newcommand{\transp}{{\sf T}}

\newcommand{\rect}{{\sf rect}}

\newcommand{\Nslot}{N_{\rm slot}}

\newcommand{\Na}{N_{\rm a}}

\newcommand{\Tcp}{T_{\rm cp}}

\usepackage[bookmarks=false,hidelinks,colorlinks=false,pdfborder={0 0 0}]{hyperref}
\hypersetup{draft}
\def\BibTeX{{\rm B\kern-.05em{\sc i\kern-.025em b}\kern-.08em
    T\kern-.1667em\lower.7ex\hbox{E}\kern-.125emX}}
\begin{document}

\title{Near-Field Integrated Sensing and Communication with a Uniform Circular Array}
\author{\IEEEauthorblockN{
	        Lorenzo Zaniboni\IEEEauthorrefmark{1},
			Mark F. Flanagan\IEEEauthorrefmark{2},
            Gerhard Kramer\IEEEauthorrefmark{1}}
		    \IEEEauthorblockA{\\
		    \IEEEauthorrefmark{1}Institute for Communication Engineering, Technical University of Munich, Germany\\
			\IEEEauthorrefmark{2}School of Electrical and Electronic Engineering, University College Dublin, Ireland\\
			Email: lorenzo.zaniboni@tum.de, mark.flanagan@ieee.org, gerhard.kramer@tum.de}}

\maketitle
\thispagestyle{plain}
\pagestyle{plain}

\begin{acronym}
	\acro{AoA}{angle of arrival}
	\acro{AoD}{angle of departure}
	\acro{AWGN}{additive white Gaussian noise}
    \acro{BS}{base station}
    \acro{BA}{beam alignment}
    \acro{BF}{beamforming}
    \acro{CRLB}{Cram\'er-Rao lower bound}
    \acro{CP}{cyclic prefix}
    \acro{DD}{delay-Doppler}
    \acro{DFT}{discrete Fourier transform}
    \acro{DL}{downlink}    
    \acro{ELAA}{extremely large-scale antenna array}
    \acro{FF}{far-field}
    \acro{FIM}{Fisher information matrix}
    \acro{FT}{Fourier transform}
    \acro{HDA}{hybrid digital-analog}
    \acro{HIRS}{hybrid-intelligent reflective surface}
    \acro{i.i.d.}{independent and identically distributed}
    \acro{IRS}{intelligent reflecting surface}
    \acro{ICI}{inter-carrier interference}
    \acro{ISI}{inter-symbol interference}
    \acro{ISAC}{integrated sensing and communication}
    \acro{LOS}{line-of-sight}
    \acro{ML}{maximum likelihood}
    \acro{mmWave}{millimeter wave}
    \acro{MIMO}{multiple-input multiple-output}
    \acro{MMLE}{multi-slot maximum likelihood estimation}
    \acro{NF}{near-field}
    \acro{OFDM}{orthogonal frequency-division multiplexing }
    \acro{RCS}{radar cross-section}
    \acro{Rx}{receiver}
    \acro{RF}{radio frequency}
    \acro{RMSE}{root mean square error}
    \acro{SNR}{signal-to-noise ratio}
    \acro{TD}{time-delay}
    \acro{Tx}{transmitter }
	\acro{ULA}{uniform linear array}
    \acro{UL}{uplink}
	\acro{UE}{user equipment}
    \acro{UCA}{uniform circular array}
    \acro{QPSK}{quadrature phase-shift keying}
\end{acronym}

\begin{abstract}
Near-field (NF) integrated sensing and communication (ISAC) is studied for a uniform circular array (UCA) and monostatic sensing. A continuous-time channel model is derived for the spherical wavefront and orthogonal frequency-division multiplexing. A multi-slot Cram\'er-Rao lower bound (CRLB) for joint range-angle-Doppler estimation is derived under a slow-motion approximation, showing how the signal delay and wavefront curvature affect the range bound. A transmit beamformer that minimizes the position CRLB is designed via gradient descent, and a maximum-likelihood estimator is obtained by using the Levenberg--Marquardt algorithm initialized with a grid search. Simulations show that increasing the UCA radius improves angle and motion estimation at all distances, but improves the range bound only at distances far inside the Rayleigh distance, and lowers the reliable communication rate.
\end{abstract}


%%%%%%%%%%%%%%%%%%%%%%%%%%%%%%%%
\section{Introduction}\label{sec:introduction}
\Ac{ISAC} is an enabling technology for 6G networks that can reduce spectrum and hardware requirements \cite{9705498,9737357}. The use of \acp{ELAA} and \ac{mmWave} frequencies in future 6G deployments pushes wireless systems into the \ac{NF}, where the spherical wavefront structure of the propagating signal should be accounted for \cite{10220205, cong2024near}. \ac{NF} measurements permit joint range-angle estimation from a single \ac{BS} \cite{wang2023near}, unlike the \ac{FF} case. This makes the \ac{NF} attractive for \ac{ISAC}.

\Acp{ULA} \cite{wang2023near,giovannetti2025asymptotic} and discrete channel models \cite{zhang2025fair} are certainly useful, but they discard two effects that the \ac{NF} makes available, which are the angle-delay correlation and the variation of the Doppler shift across the aperture \cite{wang2024rethinking}, which is what separates the speed from the direction of motion. Furthermore, \ac{ULA}-based \ac{NF} analysis reveals a limitation: the effective array aperture, and consequently the \ac{NF} region, shrinks rapidly as the target moves away from broadside, leaving large parts of the coverage area without \ac{NF} benefits \cite{wu2023enabling}. The paper \cite{giovannetti2025asymptotic} further shows that if the \ac{ULA} aperture is sufficiently large to enter the \ac{NF} regime, the \ac{CRLB} for joint range-angle estimation saturates. The saturation stems from nonlinearities in the phase profile and non-uniform path loss across elements. These limitations motivate rethinking the array geometry. 

The \ac{UCA} is an alternative: its rotational symmetry provides a nearly angle-invariant \ac{NF} region. The paper \cite{wu2023enabling} shows that the \ac{UCA}'s effective Rayleigh distance is uniform across directions, while that of the \ac{ULA} collapses. The angle-symmetric coverage is a decisive advantage for monostatic radar sensing. For instance, with $\Na=64$ elements at 60 GHz, a \ac{UCA} with radius $R=0.5$~m achieves a \ac{NF} region of $\approx 400$ m, compared to only 10 m for a \ac{ULA} with half-wavelength spacing, a 40× advantage with similar hardware. The paper \cite{xue2025impactuniformcirculararray} proposed a \ac{UCA}-based \ac{NF}-\ac{ISAC} framework, derived \acp{CRLB}, and designed an optimal beamformer under a narrowband model neglecting delay and Doppler shifts. 

Our contributions are as follows.
\begin{itemize}
    \item We combine wideband \ac{OFDM}, spherical wavefront geometry, beamforming and \ac{ML} estimation in one framework. The per-element delay and Doppler shift render the position and motion jointly identifiable.
    \item A wideband \ac{NF}-\ac{ISAC} model for a \ac{UCA}-equipped \ac{BS} and one \ac{UE} that incorporates delay, Doppler shift, and spherical wavefront propagation. The model includes \ac{OFDM} through a slow-motion approximation.
    \item \ac{CRLB} equations for joint estimation of the range, angle, velocity, and angle of motion.
    \item Transmit beamformer design via gradient descent and an \ac{ML} estimator via the Levenberg–Marquardt algorithm initialized by a grid search.
    \item Simulations reveal how the \ac{UCA} radius and wavefront curvature affect parameter estimation.
\end{itemize}
\subsubsection*{Notation} Lowercase bold letters denote vectors, and uppercase bold letters denote matrices. $(\cdot)^*$, $(\cdot)^\transp$, and $(\cdot)^\herm$ denote the complex conjugate, transpose, and Hermitian operators, respectively. $\|\xv\|$ is the $\ell_2$-norm of $\xv$. $\diag(\xv)$ is a diagonal matrix with the elements of $\xv$ on its diagonal. $\nabla$ and $\nabla^{\bot}$ are the gradient of a function and its projection, respectively. 
%$1(.)$ is the indicator function that takes on the value 1 if its argument is true, and is otherwise 0.

%%%%%%%%%%%%%%%%%%%%%%%%%%%%%%%%%%
\section{System Model} \label{sec:system}
\begin{figure}[t]
    \centering
  	\includegraphics[width=.45\textwidth]{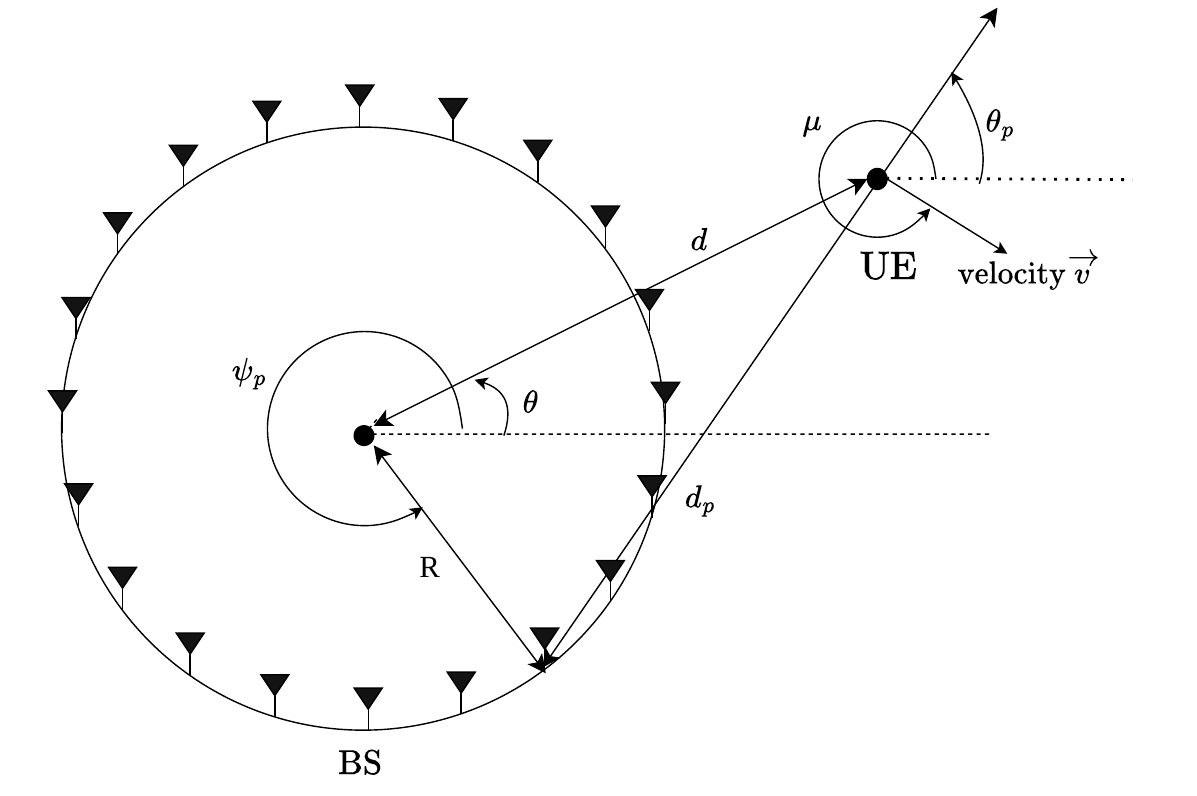}
  	\caption{System model. The origin is the center of the \ac{UCA}, so the coordinates of the \ac{UE} and antenna $p$ are $(d\cos\theta,d\sin\theta)$ and $(R\cos\psi_p,R\sin\psi_p)$, respectively.}
  	\label{fig:system_schematic}
\end{figure}

Fig.~\ref{fig:system_schematic} shows a \ac{BS} with a \ac{UCA} of radius $R$ and $N_a$ antenna elements in the azimuth plane. The \ac{UCA} center is taken to be the origin. The $p$-th element, $p=1,\dots,N_a$, is located at angle $\psi_p = 2\pi(p-1)/\Na$ relative to the reference direction, which is towards the right in the figure.

The \ac{UE} is located at position $(d \cos\theta, d \sin\theta)$, and it moves at $v$ m/s at the angle $\mu$ relative to the reference direction. The distance from the $p$-th antenna element to the \ac{UE} is
\begin{align} \label{eq:UCA_distance}
    d_p = \sqrt{d^2 + R^2 - 2dR\cos(\phi_p)}
\end{align}
where $\phi_p = \theta - \psi_p$. The angle of the line from element $p$ to the \ac{UE} relative to the reference direction is
\begin{align} \label{eq:theta_p}
    \theta_p = \arccos\left( \frac{d\cos\theta - R\cos\psi_p}{d_p} \right)
\end{align}
if the \ac{UE} lies above element $p$, as in Fig.~\ref{fig:system_schematic}. If the \ac{UE} lies below element $p$, one must add $\pi$ to the right-hand side of \eqref{eq:theta_p}. 

We wish to estimate the position $(d,\theta)$ and motion $(v,\mu)$ parameters. We collect these in the vector $\etav = [d,\theta,v,\mu]^\transp$. These parameters are time-varying in general; we omit their time dependence for notational convenience.

%%%%%%%%%%%%%%%%%%%%%%%%%%%%
\subsection{Channel Model}
Consider the carrier frequency $f_c$ and wavelength $\lambda=c/f_c$, where $c$ is the speed of light. Suppose the \ac{UE} is located within the \ac{NF} region of the \ac{BS}, meaning that $d$ is less than the Rayleigh distance $\frac{2 (2R)^2}{\lambda}$. The channels are characterized by $\Na$ paths with delays $\tau_p = d_p/c$, gains $g_p$, and Doppler shifts $\nu_p$. The free-space gains are collected in the vector
\begin{align*} %\label{eq:gain}
    \gv = \sqrt{G_{\rm BS} G_{\rm UE}}\, 
    \frac{\lambda}{4\pi}
    \big[ 1/d_1,\, \dots,\, 1/d_{N_a} \big]^\transp   
\end{align*}
where $G_{\rm UE},G_{\rm BS}$ are antenna gains that we normalize to 1. The \ac{UCA} steering vector may be defined as (see \cite[Eq. (7)]{10220205})
\begin{align} \label{eq:UCA_steering_array}
        \bar{\av} = \frac{e^{j2\pi d/\lambda}}{\sqrt{\Na}} \Big[ e^{-j \frac{2\pi}{\lambda} d_1}, \dots, e^{-j \frac{2\pi}{\lambda} d_{\Na}} \Big]^\transp 
\end{align}
which captures the array phase response at position $(d,\theta)$ and wavelength $\lambda$. The term $e^{j2\pi d/\lambda}$ in \eqref{eq:UCA_steering_array} normalizes the distances and is removed by the channel. We will instead work with the unnormalized steering vector
\begin{align} \label{eq:UCA_steering_array2}
        \av = \frac{1}{\sqrt{\Na}} \Big[ e^{-j \frac{2\pi}{\lambda} d_1}, \dots, e^{-j \frac{2\pi}{\lambda} d_{\Na}} \Big]^\transp .
\end{align}
The Doppler shifts are
\begin{align}
    \nu_p & = -\frac{f_c\, v}{c}\, \cos (\mu-\theta_p) \nonumber \\
    & = -\frac{f_c\, v}{c}\, \big[
    \cos\mu\,\cos\theta_p + \sin\mu\,\sin\theta_p 
    \big] .
    \label{eq:UCA_Doppler}
\end{align}
%where (see~\eqref{eq:theta_p})
%\begin{align*}
%    \cos\theta_p & = \frac{d\cos\theta - R\cos\psi_p}{d_p}, \quad
%    \sin\theta_p = \frac{d\sin\theta + R\sin\psi_p}{d_p}.
%\end{align*}
The \ac{BS}-to-\ac{UE} channel is
\begin{align} \label{eq:comm_channel}
    \hv_{\rm comm}(t,\tauv)
    = \underbrace{(\gv \odot \av \odot e^{j\pi\nuv t})}_{\displaystyle \hv(t)}
    \odot\, \delta(\tau - \tauv) 
\end{align}
where $e^{j\pi\nuv t}$ and $\delta(\tau - \tauv)$ are $N_a \times 1$ vectors with entries $e^{j\pi\nu_p t}$ and $\delta(\tau - \tau_p)$ at position $p$, respectively.

The \ac{UE} is modeled as a point target that reflects the incident waveform back toward the \ac{BS}. The round-trip sensing channel is thus the symmetric $N_a \times N_a$ matrix
\begin{align*} %\label{eq:sens_channel}
    \Hm_{\rm sens}(t,\Tm) = \rho\, 
    \big( \hv(t)\, \hv(t)^\transp \big)
    \odot \delta(\tau - \Tm)
\end{align*}
where $\rho$ is the complex reflectivity representing the \ac{RCS} and $\Tm$ is the symmetric $N_a \times N_a$ matrix with $p,q$ entry $\tau_p+\tau_q$ for $1\le p,q\le N_a$.

%%%%%%%%%%%%%%%%%%%%%%%%%%%%
\subsection{OFDM Signaling}
Consider $\Nslot$ slots each having $N$ \ac{OFDM} symbols with $M$ subcarriers. Let $x_i[n,m]$ be the transmit symbol of \ac{OFDM} symbol $n$ and subcarrier $m$. The $i$-th slot signal is
\begin{align*} %\label{eq:OFDM_signal}
    \sv_i(t) = \fv_i \sum_{n=0}^{N-1} \sum_{m=0}^{M-1} 
    x_i[n,m]\, & \rect \left( \frac{t - nT_o}{T_o} \right)
    \nonumber \\
    & \cdot e^{j 2 \pi m f_s (t - nT_o - \Tcp)}
\end{align*}
where
$\fv_i$ is the \ac{BF} vector of slot $i$, $\rect(x)=1(0\le x < 1)$ is a rectangular pulse,
$f_s$ is the subcarrier spacing,
$\Tcp$ is the \ac{CP} duration, and
$T_o$ is the \ac{OFDM} symbol time including $\Tcp$. 
The \ac{BF} vectors $\fv_i$ have unit norm, and the transmit symbols satisfy the power constraint
\begin{align*} %\label{eq:power_constraint}
    \frac{1}{M} \sum\nolimits_{m} \EE [|x_i[n,m]|^2] 
    \le P_t, \quad \forall(i,n,m)
\end{align*}
where $\EE[.]$ denotes expectation. We assume $\tau_{\max}=\max_p \tau_p$ and $\nu_{\max}=\max_p \nu_p$ satisfy
\begin{align} \label{eq:delay_subcarrier_constraints}
    2\, \tau_{\rm max} \le \Tcp \ll T_o, \quad
    2\, \nu_{\rm max} \ll f_s
\end{align}
where the factor 2 emphasizes that the relations should be valid for both the communication and sensing channels.

%%%%%%%%%%%%%%%%%%%%%%%%%%%%%%%%%%%
\subsection{UE Signal}
The \ac{UE} received signal is a sum of convolutions:
\begin{align*}
    y_i(t) = w_i(t) & + \sum\nolimits_p\ \int_{-\infty}^\infty
    [\hv_{\rm comm}(t,\tauv)]_p\,
    [\sv_i]_p(t-\tau)\, d\tau
    \nonumber \\
    = w_i(t) & + \sum_{p,n,m} x_i[n,m]\,
    \rect \left( \frac{t - \tau_p - nT_o}{T_o} \right)
    \nonumber \\
    &\qquad \cdot [\hv(t)]_p\, [\fv_i]_p\, 
    e^{j 2 \pi m f_s(t - \tau_p - nT_o - \Tcp)}
\end{align*}
where $w_i(t)$ is \ac{AWGN}. The \ac{UE} computes $MN$ projections $y_i[n,m]$ for all $m,n$:
\begin{align} \label{eq:UE-projections}
    \int_{nT_o+\Tcp}^{(n+1)T_o} y_i(t)\, 
    e^{-j 2 \pi m f_s(t - nT_o)}\,
    \rect \left( \frac{t - nT_o}{T_o} \right)\, dt .
\end{align}

Evaluating \eqref{eq:UE-projections} usually involves approximations. For example, we interpret the first assumption in \eqref{eq:delay_subcarrier_constraints} as giving
\begin{align} \label{eq:delay-approx}
    \rect \left( \frac{t - \tau_p - nT_o}{T_o} \right)
    \approx \rect \left( \frac{t - nT_o}{T_o} \right).
\end{align}
Also, one typically assumes that the vehicle moves slowly relative to the chip time $(T_o-\Tcp)/M$. However, the vehicle may not be moving slowly relative to the \ac{OFDM} symbol time $T_o$, and this leads to \ac{ICI}. We ignore the \ac{ICI} and interpret the second assumption in \eqref{eq:delay_subcarrier_constraints} as giving
\begin{align} \label{eq:Doppler-approx}
    e^{j \pi \nu_p t}
    \approx e^{j \pi \nu_p n T_o}, \quad
    nT_o\le t\le (n+1)T_o.
\end{align}
This approximation is accurate for slow \acp{UE}, and it simplifies processing by modeling the subcarriers as orthogonal.

Applying \eqref{eq:delay-approx}-\eqref{eq:Doppler-approx}, the receiver uses the model
\begin{align} \label{eq:UE-output}
    y_i[n,m] = w_i[n,m] 
    + & x_i[n,m]\, e^{-j2\pi m f_s \Tcp} 
    \nonumber \\
    & \cdot \sum\nolimits_p [\hv[n,m]]_p [\fv_i]_p
\end{align}
where $w_i[n,m] \sim \mathcal{CN}(0,\sigma^2)$ is \ac{i.i.d.} across $i,m,n$, and where
\begin{align} \label{eq:hvnmp}
    [\hv[n,m]]_p 
    = [\hv(n T_o)]_p\, e^{-j 2 \pi m f_s \tau_p} .
%    \\
%    & = g_p \frac{1}{\sqrt{\Na}} e^{j\frac{2\pi}{\lambda}(d-d_p)}
%    e^{-j 2 \pi m f_s \tau_p}
%    e^{j \pi \nu_p n T_o}.
\end{align}
We thus consider the parallel channels
\begin{align} \label{eq:UE-channel}
    y_i[n,m] = \beta_i[n,m]\, x_i[n,m] + w_i[n,m]
\end{align}
where the channel coefficients are
\begin{align*}
    \beta_i[n,m] = \hv[n,m]^\transp\, \fv_i\, 
    e^{-j2\pi m f_s \Tcp} .
\end{align*}

%%%%%%%%%%%%%%%%%%%%%%%%%%%%%%%%%%%
\subsection{BS Signal}
The back-scattered signal at the \ac{BS} is modeled as
\begin{align*}
     \yv_i[n,m] = \tilde{\wv}_i[n,m] 
     + & \rho\, x_i[n,m]\, e^{-j2\pi mf_s \Tcp}
     \nonumber \\
     &\cdot \hv[n,m] \sum\nolimits_q [\hv[n,m]]_q [\fv_i]_q
     %\label{eq:BS-output}
\end{align*}
where $\tilde{\wv}_i[n,m] \sim \mathcal{CN}(0,\sigma^2 \Id)$ is \ac{i.i.d.} across $i,n,m$.
We thus consider the parallel channels
\begin{align}\label{eq:BS-channel}
    \yv_i[n,m] = 
    \big(\rho\, \beta_i[n,m]\, \hv[n,m]\big)\, x_i[n,m] 
     + \tilde{\wv}_i[n,m]
\end{align}
and proceed to analyze the sensing performance.

%%%%%%%%%%%%%%%%%%%%%%%%%%%%%%%%
\section{Optimization and Estimation Analysis}
\subsection{Cram\'er--Rao Lower Bound}\label{sec:CRLB}
We derive the \ac{CRLB} for estimating $\etav = [d, \theta, v, \mu]^\transp$ in time slot $i$. The noise-free output in \eqref{eq:BS-channel} is
\begin{align} \label{eq:noise-free}
    \muv_i(\etav) = \rho\, \beta_i[n,m]\, \hv[n,m]\, x_i[n,m]
\end{align}
Under the \ac{AWGN} model, the per-slot $4\times 4$ \ac{FIM} has entries
%{\color{red} [There should be an expectation here.]}
\begin{align} \label{eq:FIM}
    [\Jm_i(\etav)]_{a,b}
    = \frac{2}{\sigma^2}\Real\left\{ 
    \sum\nolimits_{m,n} \EE\left[\frac{\partial \muv_i^\herm}{\partial a}
    \frac{\partial \muv_i}{\partial b}\right] \right\} 
\end{align}
for $a,b\in\{d,\theta,v, \mu\}$.
From \eqref{eq:UCA_distance} and \eqref{eq:UCA_Doppler}, we have the derivatives
\begin{align*}
    \frac{\partial d_p}{\partial \etav} 
    & = \frac{1}{d_p} \begin{bmatrix}
        d - R\cos\phi_p &
        dR\sin\phi_p & 0 & 0
    \end{bmatrix}^\transp
\end{align*}
and
\begin{align*}
    \frac{\partial \nu_p}{\partial \etav} 
    & = -\frac{f_c\, v}{c\, d_p}\! 
    \left(\!\begin{bmatrix}
        \cos(\mu-\theta) \\[1pt]
        d\sin(\mu-\theta) \\[1pt]
        (d_p/v) \cos(\mu-\theta_p) \\[1pt]
        - d_p \sin(\mu-\theta_p)
    \end{bmatrix}\!
    - \cos(\mu-\theta_p)
    \frac{\partial d_p}{\partial \etav}\! \right)\!.
\end{align*}

The expressions \eqref{eq:comm_channel} and \eqref{eq:hvnmp} give
\begin{align*}
    \left[\hv[n,m]\right]_p 
    = \frac{g_p}{\sqrt{\Na}} \, e^{-j \frac{2\pi}{\lambda} d_p}\, e^{-j 2\pi mf_s d_p/c}\,
    e^{j 2\pi \nu_p n T_o} .
\end{align*}
Applying the product rule for derivatives, we obtain
\begin{align} \label{eq:h_deriv}
    \frac{\partial [\hv[n,m]]_p}{\partial \eta} 
    = [\hv[n,m]]_p \, \uv_p[n,m]
\end{align}
where, using $f_m = f_c + m f_s$, the vector $\uv_p[n,m]$ is
\begin{align*}
    \begin{bmatrix}
        u_p^d[n,m] \\[2pt] 
        u_p^\theta[n,m] \\[2pt] 
        u_p^v[n,m] \\[2pt] 
        u_p^\mu[n,m]
    \end{bmatrix}
    = 
    - \underbrace{\frac{1}{d_p} \frac{\partial d_p}{\partial \etav}}_\text{gain}
    - \underbrace{j \frac{2\pi f_m}{c} \frac{\partial d_p}{\partial \etav}}_\text{delay}
    + \underbrace{j 2 \pi n T_o \frac{\partial \nu_p}{\partial \etav}}_{\text{Doppler}}.
\end{align*}
The product rule applied to $[\muv_i]_p$ in \eqref{eq:noise-free} yields
\begin{align} \label{eq:mu_deriv}
    \frac{\partial [\muv_i]_p}{\partial \eta}
    = \rho\, [\hv[n,m]]_p\, \Gamma_{i,p}^\eta[n,m] 
\end{align}
where $\eta\in\{d,\theta,v,\mu\}$ and $\Gamma_{i,p}^\eta[n,m]$ is
\begin{align}
    x_i[n,m] \left(u_p^\eta[n,m]\, \beta_i[n,m] + \frac{\partial \beta_i[n,m]}{\partial \eta}\right)
    \label{eq:Gamma_ip}
\end{align}
with
\begin{align*}
    \frac{\partial \beta_i[n,m]}{\partial \eta}
    = \left( \hv[n,m]^\transp \Dm^\eta \fv_i \right)\, e^{-j2\pi m f_s \Tcp}
\end{align*}
and $\Dm^\eta[n,m] = \diag\left(u_1^\eta[n,m],\dots,u_{\Na}^\eta[n,m]\right)$. The first term in the bracket in \eqref{eq:Gamma_ip} is the receive sensitivity and the second is the transmit one.  
% The motion parameters enter through the Doppler phase, and $\Gamma_{i,p}^v$ and $\Gamma_{i,p}^\mu$ factor as
% {\color{red} [Check if there should be no negative for the second part. But do we need the next four equations? They are all already in (19).]}
% \begin{align*}
%     \Gamma_{i,p}^v[n,m] &= - j 2 \pi n T_o\, \frac{f_c}{c}\, \left(\cos(\mu-\theta_p)\beta_i[n,m] + \zeta_i^v[n,m]\right) \\
%     \Gamma_{i,p}^\mu[n,m] &= - j 2 \pi n T_o\, \frac{f_c v}{c} \left(\sin(\mu -\theta_p) \beta_i[n,m] + \zeta_i^\mu[n,m]\right)
% \end{align*}
% with {\color{red} [Check here also.]}
% \begin{align*}
%     \zeta_i^v[n,m] &= e^{-j2\pi m f_s \Tcp}
%     \sum\nolimits_p [\hv[n,m]]_p\, [\fv_i]_p \cos(\mu-\theta_p) \\
%     \zeta_i^\mu[n,m] &= e^{-j2\pi m f_s \Tcp} 
%     \sum\nolimits_p [\hv[n,m]]_p\, [\fv_i]_p \sin(\mu-\theta_p).
% \end{align*}
Substituting \eqref{eq:mu_deriv} into \eqref{eq:FIM} and using $\left|[\hv[n,m]]_p\right|^2= \lambda^2 / (16\pi^2 d_p^2 \Na)$ gives
\begin{align}
    [\Jm_i(\etav)]_{a,b} 
    & = K\, \Real \left\{\sum_{m,n}\sum_p \frac{\left(\Gamma_{i,p}^a[n,m]\right)^{\ast}\Gamma_{i,p}^b[n,m]}{d^2_p}\right\}
\end{align}
where $K = |\rho|^2 \lambda^2 / (8 \pi^2 \sigma^2 \Na)$. 
%The multi-slot \ac{FIM} is $\Jm = \sum_i \Jm_i$. 

%%%%%%%%%%%%%%%%%%%%%%%%%%%%%%%
\subsection{Beamformer Design}\label{sec:BF}
Let $\Cm(\etav,\fv_i)$ be the inverse of the $(d,\theta)$ block of $\Jm_i$,  i.e. the \ac{CRLB} on $(d,\theta)$ for fixed $(v,\mu)$. We do not use the $(v,\mu)$ block because $\Jm_i$ is singular at $v=0$. The motion is treated as known for beam design, while Sec.~\ref{sec:ML} describes the \ac{ML} estimator for all four parameters to compare with the \ac{CRLB}. The \ac{FIM}~\eqref{eq:FIM} depends on $\fv_i$ through $\beta_i$ and $\partial\beta_i/\partial\eta$ in~\eqref{eq:Gamma_ip}.

For estimation, we assume that the pair $(v, \mu)$ is constant over the current collection of $\Nslot$ slots, and that its value is known from the previous slots. We then choose the beam that minimizes the position's \ac{CRLB}:
\begin{align} \label{eq:CRLB_optimization}
	\hat{\fv}_i = \argmin_{\fv_i:\: \|\fv_i\|=1} \Tr(\Cm(\etav,\fv_i))
\end{align}
where \eqref{eq:CRLB_optimization} uses the estimated position from the previous slot. The objective is
\begin{align}\label{eq:obj_fun_optim}
    \Tr(\Cm(\etav,\fv_i)) = \frac{[\Jm_i]_{d,d} + [\Jm_i]_{\theta,\theta}}{[\Jm_i]_{d,d}[\Jm_i]_{\theta,\theta} - [\Jm_i]_{d,\theta}^2}
\end{align}
which we minimize by gradient descent. Define
\begin{align*}
    \mathcal{R} = [\Jm_i]_{d,d}, \quad
    \mathcal{T} = [\Jm_i]_{\theta,\theta}, \quad \mathcal{X} = [\Jm_i]_{d,\theta} .
\end{align*}
Since $\beta_i$ is linear in $\fv_i$, we have the Wirtinger derivative
\begin{align*}
    \frac{\partial \Gamma_{i,p}^\eta[n,m]}{\partial \fv_i}
    & = x_i[n,m]\, \hv[n,m]^\transp \left(u_p^\eta[n,m]\, \Id + \Dm^\eta \right)\\
    &\quad\cdot e^{-j2\pi m f_s \Tcp}.
\end{align*}
The Wirtinger derivatives in \eqref{eq:wirtinger} at the top of the page are
\begin{align*}
    \frac{\partial\mathcal{R}}{\partial\fv_i}
    &= \sum_{m,n}\sum_p \frac{\left(\Gamma_{i,p}^{d}\right)^{\ast}}{d_p^2}
       \frac{\partial \Gamma_{i,p}^{d}}{\partial \fv_i} \\
    \qquad
    \frac{\partial\mathcal{T}}{\partial\fv_i}
    &= \sum_{m,n}\sum_p \frac{\left(\Gamma_{i,p}^{\theta}\right)^{\ast}}{d_p^2}
       \frac{\partial \Gamma_{i,p}^{\theta}}{\partial \fv_i} \\
    \frac{\partial\mathcal{X}}{\partial\fv_i}
    &= \frac{1}{2}\sum_{m,n}\sum_p \frac{1}{d_p^2}
    \left[
      \left(\Gamma_{i,p}^{d}\right)^{\ast}\frac{\partial \Gamma_{i,p}^{\theta}}{\partial \fv_i}
    + \left(\Gamma_{i,p}^{\theta}\right)^{\ast}\frac{\partial \Gamma_{i,p}^{d}}{\partial \fv_i}
    \right] .
\end{align*}

\newcounter{TempEqB}\setcounter{TempEqB}{\value{equation}}
\begin{figure*}[t!]\normalsize
\begin{align} \label{eq:wirtinger}
    \frac{\partial\Tr(\Cm)}{\partial \fv_i} 
    = \frac{(\mathcal{RT} -\mathcal{X}^2)\bigl(\frac{\partial\mathcal{R}}{\partial \fv_i}
    + \frac{\partial\mathcal{T}}{\partial \fv_i}\bigr) - (\mathcal{R}
    + \mathcal{T})\bigl(\mathcal{T}\frac{\partial\mathcal{R}}{\partial \fv_i}
    + \mathcal{R}\frac{\partial\mathcal{T}}{\partial \fv_i}
    - 2\mathcal{X}\frac{\partial\mathcal{X}}{\partial \fv_i}\bigr)}{(\mathcal{RT}-\mathcal{X}^2)^2}
\end{align}
\hrulefill\vspace*{4pt}
\end{figure*}
\setcounter{equation}{\value{TempEqB}}\stepcounter{equation}

The Riemannian gradient on the unit sphere is the projection of the Euclidean gradient onto the tangent space at $\fv_i^{(t)}$:
\begin{align*}
    \nabla^\perp_{\fv_i}\Tr(\Cm) = \left(\frac{\partial\Tr(\Cm)}{\partial\fv_i}\right)^\herm - \Real\left\{\frac{\partial\Tr(\Cm)}{\partial\fv_i}\left(\fv_i^{(t)}\right)^\herm\right\}\fv_i^{(t)}
\end{align*}
where $\Tr(\Cm)$ is real and $\fv_i^{(t)}$ is the $t$-th iterate \cite{absil2008optimization, brandwood1983complex, kreutz2009complex}. The iterates follow the negative projected gradient with step parameter $\alpha_t$ chosen by Armijo backtracking~\cite[Cor.~2]{armijo1966minimization}, \cite[Alg.~3.1]{nocedal2006numerical}. They are then renormalized, and the process terminates when $\|\nabla^\perp_{\fv_i}\Tr(\Cm)\|<\varepsilon$. The per-slot objective depends only on the position, so the beam is recomputed in every slot. Sec.~\ref{sec:sim} compares beams built from the true position and from the \ac{ML} estimate.

%%%%%%%%%%%%%%%%%%%%%%%%%%%%%%%
\subsection{ML Estimation of $\etav=[d,\theta,v,\mu]^\transp$}\label{sec:ML}
The \ac{ML} estimator collects the $\{\hat{\fv}_i\}$ from Sec.~\ref{sec:BF}. Let $\Bar{\yv}$ and $\Bar{\muv}(\etav)$ be the stacked vectors from the $\Nslot$ slots. We have
\begin{align*}
    \Bar{\yv} = \Bar{\muv}(\etav) + \nv
\end{align*}
with $\nv\sim\mathcal{CN}(0,\sigma^2\Id)$. Maximizing the log-likelihood minimizes $L(\etav) = \|\bar{\yv} - \bar{\muv}\|^2$, so for $\eta\in\{d,\theta,v,\mu\}$ we set
\begin{equation}\label{eq:derivative_ML}
    \Real\left\{(\partial\Bar{\muv}^\herm/\partial\eta)
    (\Bar{\yv}-\Bar{\muv})\right\} = 0 .
\end{equation}
Substituting \eqref{eq:noise-free} and \eqref{eq:mu_deriv} into \eqref{eq:derivative_ML}, we obtain
\begin{align}\label{eq:score}
    F_\eta(\etav) = \Real\Bigg\{
    \sum\nolimits_{i,n,m,p} \left(\Gamma_{i,p}^\eta\right)^\ast \varrho_{i,p}
    \Bigg\} = 0
\end{align}
where
\begin{align*}
    \varrho_{i,p} =
    \rho^\ast\,
    [\hv[n,m]]_p^\ast \, [\Bar{\yv}_i[n,m]]_p
    - \frac{K\sigma^2}{2\, d_p^2} x_i[n,m]\,\beta_i .
\end{align*}
%The scores \eqref{eq:score} expand as in \eqref{eq:ML_score} on the next page.
%
%\newcounter{TempEqCnt3}\setcounter{TempEqCnt3}{\value{equation}}
%\begin{figure*}[!t]\normalsize
%\begin{align} \label{eq:ML_score}
%    F_\eta(\etav) = \sum_{i,n,m,p}\left[\Real\{u_p^\eta\}\,\Real\{x_i^\ast\beta_i^\ast\varrho_{i,p}\} + \Imag\{u_p^\eta\}\,\Imag\{x_i^\ast\beta_i^\ast\varrho_{i,p}\} + \Real\bigl\{x_i^\ast(\partial \beta_i/\partial \eta)^\ast\varrho_{i,p}\bigr\}\right] = 0
%\end{align}
%\hrulefill\vspace*{4pt}
%\end{figure*}
%\setcounter{equation}{\value{TempEqCnt3}}\stepcounter{equation}

The cost $L(\etav)$ is non-convex with multiple local minima in the azimuth plane. This is because the element spacing $2\pi R/\Na\approx10\lambda$ at $R=0.5$\,m causes the transmit beam pattern to have strong sidelobes at angles near multiples of $2\pi/\Na$. We thus initialize with a coarse grid search over $(d,\theta)$, with a step in $d$ that follows the \ac{NF} focusing width, then initialize $(v, \mu)$ by a coarse Doppler search, and refine all four parameters with the Levenberg--Marquardt algorithm.

%%%%%%%%%%%%%%%%%%%%%%%
\section{Simulations}\label{sec:sim}
\begin{table}[b]
  \centering
  \caption{System parameters.}
  \label{tab:params}
  \renewcommand{\arraystretch}{1.1}
  \setlength{\tabcolsep}{4pt}
  \begin{tabular}{@{}ll@{\hspace{30pt}}ll@{}}
    \toprule
    $R$ & $\frac14,\,\frac12,\,1,\,2$\,m & $M\Pt$ & $30$\,dBm \\
    $\Na$ & $64$ & Noise figure & $10$\,dB \\
    $f_c$ & $60$\,GHz & RCS & $-10$\,dBsm \\
    $M$, $f_s$ & $1024$, $480$\,kHz & $d$ & $5,7,\dots,21$\,m \\
    $\Tcp$ & $146$\,ns & $\theta$, $\mu$ & $30^\circ$, $45^\circ$ \\
    $N$, $\Nslot$ & $14$, $200$ & $v$ & $5$\,m/s \\
    \bottomrule
  \end{tabular}
\end{table}

The simulation parameters are shown in Table~\ref{tab:params}. We used \ac{QPSK} pilots, bandwidth $B=Mf_s\approx 492$\,MHz, and a \ac{RCS} with $|\rho|^2=4\pi\sigma_{\mathrm{RCS}}/\lambda^2$ and $\lambda=5$\,mm. The \ac{UE} follows a constant-velocity trajectory and $\etav=[d,\theta,v,\mu]^\transp$ is taken at the midpoint of $\Nslot$ slots, over which the \ac{UE} moves $\Nslot\,N\, (1/f_s+\Tcp)\cdot v \approx 31$ mm. A single $\etav$ is estimated from all $\Nslot$ slots.
All distances are less than the Rayleigh distance, which is at least $100$\,m for all cases. The received signal is generated using \eqref{eq:UE-projections} without the approximations \eqref{eq:delay-approx}--\eqref{eq:Doppler-approx}. The parameters $d_p$, $\tau_p$, $g_p$, and $\nu_p$ are updated in every slot, whereas the \ac{CRLB}, the beamformer, and the estimator use the slow-motion model \eqref{eq:BS-channel}. The phase $\arg\,\rho$ is drawn uniformly at random in each trial, and the receiver computes the \ac{ML} estimate of $\arg\,\rho$ based on \eqref{eq:BS-channel}--\eqref{eq:noise-free}. Each point in the figures below is based on 150 trials. We say a trial fails if the grid search returns a local minimum different than the global one; see Sec.~\ref{sec:ML}. The \ac{RMSE} is computed over the successful trials and shown only if the fraction of failures is below $10\%$.

\begin{figure}[t]
    \centering
    \includegraphics[width=0.77\columnwidth]{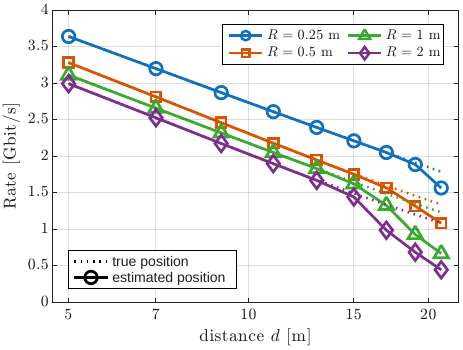}
    \caption{Achievable rate at the UE.}
    \label{fig:rate}
\end{figure}

Fig.~\ref{fig:rate} shows the achievable information rates. The rate decreases with radius $R$ from $3.6$ to $3.0$\,Gbit/s at $d=5$\,m. The figure compares the beams built from the true $\etav$ and the \ac{ML} estimate $\hat\etav$, averaged over all trials. The two coincide up to $d=13$\,m, after which they separate first for the largest radius. At $d=21$\,m, the \ac{ML} estimate loses $13\%$ of the rate for $R=0.25$\,m and $59\%$ for $R=2$\,m. The loss is caused by the failures referred to above.

\begin{figure}[t]
    \centering
    \begin{subfigure}[b]{0.77\columnwidth}
        \includegraphics[width=\linewidth]{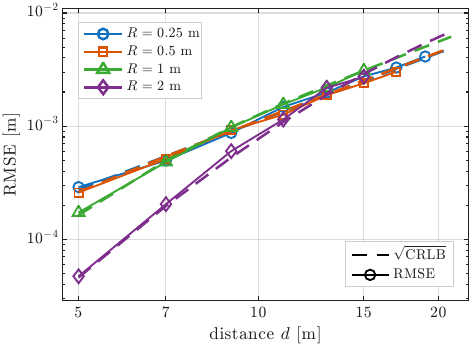}
        \caption{Range}
        \label{fig:rmse_d}
    \end{subfigure}
    \par\medskip
    %\hfill
    \begin{subfigure}[b]{0.77\columnwidth}
        \includegraphics[width=\linewidth]{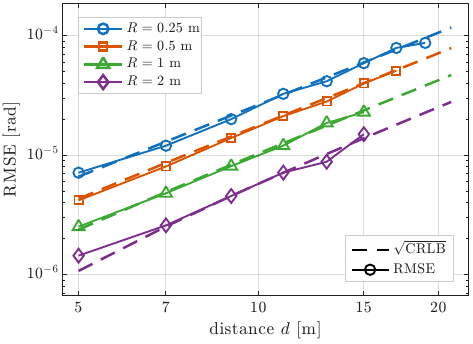}
        \caption{Angle}
        \label{fig:rmse_theta}
    \end{subfigure}
    %\caption{RMSE and CRLB of the position estimates.}
    %\label{fig:rmse_pos}
%\end{figure}
%\begin{figure}[t]
    %\centering
    \par\medskip
    \begin{subfigure}[b]{0.77\columnwidth}
        \includegraphics[width=\linewidth]{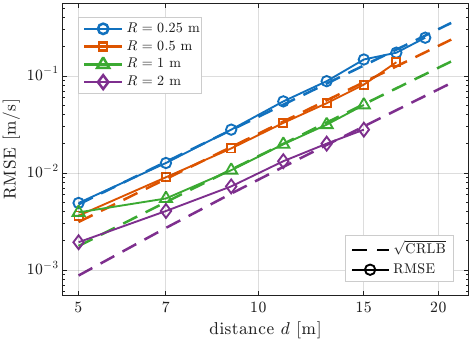}
        \caption{Velocity}
        \label{fig:rmse_v}
    \end{subfigure}
    \par\medskip
    %\hfill
    \begin{subfigure}[b]{0.77\columnwidth}
        \includegraphics[width=\linewidth]{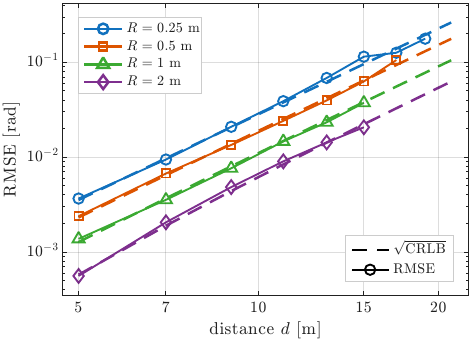}
        \caption{Direction of motion}
        \label{fig:rmse_mu}
    \end{subfigure}
    \caption{RMSE and CRLB of the estimates.}
    \label{fig:rmse_mot}
\end{figure}

Fig.~\ref{fig:rmse_mot} shows that the sensing performance improves as $R$ increases. A larger $R$ tightens the angle and motion bounds (Fig.~\ref{fig:rmse_theta}--\ref{fig:rmse_mu}) by $4$--$6\times$ at all distances. The range (Fig.~\ref{fig:rmse_d}) behaves differently: at $d=5$\,m the bound for $R=2$\,m is $6\times$ tighter than for $R=0.25$\,m, but the advantage shrinks as $d$ increases, and the curves cross near $d=13$\,m. The \ac{ML} estimator is within $20\%$ of the bounds for $d\geq9$\,m. At $d=5$\,m the velocity \ac{RMSE} (Fig.~\ref{fig:rmse_v}) for $R\geq1$\,m lies $2\times$ above the \ac{CRLB}, and the slow-motion model \eqref{eq:BS-channel} is no longer accurate. The distance beyond which more than $10\%$ of the trials fail, and the \ac{RMSE} is thus not shown, decreases from $19$\,m at $R=0.25$\,m to $15$\,m at $R=2$\,m.

%%%%%%%%%%%%%%%%%%%%%%%
\section{Conclusions}
We have developed a wideband \ac{NF}-\ac{ISAC} framework for a \ac{UCA}-equipped \ac{BS} and monostatic sensing. We derived a multi-slot \ac{CRLB} for estimating position and motion, designed transmit beamformers to minimize the position \ac{CRLB} via gradient descent, and obtained an \ac{ML} estimator using the Levenberg--Marquardt algorithm initialized with a grid search. Simulations show that increasing the \ac{UCA} radius improves angle and motion estimation at all distances, but confines the range gain to distances far inside the Rayleigh distance, and lowers the rate. The rate is negligible up to $13$\,m. The loss beyond this distance is due to estimation failures, so the rate and sensing accuracy degrade together.

\clearpage
\section*{Acknowledgments}
The work was funded by the Deutsche Forschungsgemeinschaft (DFG) through research grants {KR~3517/12-1} and {KR~3517/13-1}. 
The AI tool Claude~Version Opus 4.8 was used to assist with writing the optimization code.

\bibliographystyle{IEEEtran}
\bibliography{NFC,IEEEabrv}
\end{document}